# Evaluating the Impact of Features Selection Methods on SNMP-MIB Interface Parameters to Accurately Detect Network Anomalies

Ghazi Al-Naymat
Ajman University, UAE
g.alnaymat@ajman.ac.ae

Ahmed Hambouz
Princess Sumaya University for Technology, Jordan
ahmedhambouz@gmail.com

Mouhammd Al-kasassbeh
Princess Sumaya University for Technology, Jordan
m.alkasassbeh@psut.edu.jo

*Abstract*— **Many approaches have evolved to enhance the process of detecting network anomalies using SNMP-MIBs. Most of these approaches focus on machine learning algorithms with a lot of SNMP-MIB database parameters, which may consume most of the hardware resources (CPU, memory, and bandwidth). In this paper, we introduce an efficient detection model to detect network anomalies using Lazy.IBk as a machine learning classifier, Correlation, and ReliefF as an approach for attribute evaluators only SNMP-MIB interface parameters. This model achieves high accuracy 99.94% with minimal hardware resources consumption. Thus, this model can be adopted in the intrusion detection system (IDS) to increase its performance and efficiency.**

*Keywords—Network attacks, SNMP, anomaly detection, SNMP-MIB interface parameters.*

## I. INTRODUCTION

Most of the enterprises in the world are connected to the internet. These enterprises are actually under the risk of huge losses or termination of functionality due to the cessation of their introduced services through the internet. A major cessation cause of enterprise Internet services is the Denial of Service (DoS) attacks. DoS attack is a targeted attack on the server, router, or network to prevent it from serving the legitimate users in a normal manner [1]. In DoS attacks, the victim resources are highly consumed by the attacker to deprive legitimate users of being served [2].

### A. Network Attacks

Many network attacks can be classified as DoS attacks. Following are some examples:

- TCP-SYN flood attack: This attack exploits a weakness in TCP handshaking model. In TCP three-way handshaking model, the client sends a SYN packet to the server, the server responds with a SYN-ACK packet to the client, stores the request parameters in the memory, and waits for ACK packet. In TCP-SYN flood attack, the attacker sends SYN request packet to the router with spoofed IP address, the router responds with a SYN-ACK packet for the spoofed IP and waits for an ACK packet from the client that will never reach. The attacker repeats this process many times to exhaust the router resources. Thus, the router resources will be wasted by allocating memory spaces to fake requests.

- UDP Flood Attack: the attack occurs when the attacker sends numerous packets to random ports on the victim computer, the victim in turn, responds to these packets with the appropriate ICMP packets. The huge traffic of the attack requests and the responses will slow down the victim's computer bandwidth and resources [3].

- ICMP-ECHO Attack: In this attack, the attacker sends a huge number of ICMP requests or large ICMP packets with a spoofed IP address (victim's IP address) to many nodes in the network, the nodes respond to these requests and send the responses to the spoofed IP address, the victim. This huge traffic will slow down the victim by wasting its resources, such as bandwidth, CPU, memory.

- HTTP Flood Attack: The attacker sends a large number of HTTP requests to the victim web server. These requests are similar to legitimate requests from normal users. The attacker uses botnets to generate requests to the victim web server. This attack is considered a complicated attack because it is very difficult to distinguish between the attack traffic and the normal legitimate traffic of users.

- Slowloris Attack: In HTTP protocol, the Get message's header should be completely received before the header being processed. If the header is not completely received the web server will wait for the rest parts of the header presuming that the client has a slow internet connection. While of that, the web server will keep resources reserved for the http session. In Slowloris attack, the attacker sends a fake header slowly to the webserver to waste the server resources for a long time [4].

- Slowpost Attack: It is similar to the Slowloris attack, but with a difference that is in Slowloris attacks, the webserver gets the header slowly, and in the Slowpost attack gets the message body slowly [5].

### B. Simple Network Management Protocol (SNMP)

SNMP is a standard protocol emerged in the late 1980s [6] to manage devices supporting this protocol. Most network devices support SNMP, such as routers, switches, modems, servers, Uninterruptable Power Supplies (UPS), and more [7].

SNMP contains variables that describe the traffic passes through the devices. So, these variables can be monitored to characterize the devices and their traffic. The analysis of the trapped variable from many devices in the network can help to detect network anomalies [8]. SNMP variables can be grouped in the following groups: Interface group (IF), Internet Protocol (IP), Internet Control Message Protocol (ICMP), Transmission Control Protocol (TCP), User Datagram Protocol (UDP), Address Translation (AT), Exterior Gateway Protocol (EGP), and SNMP [9].

The Management Information Base (MIB) is a database of variables related to a device to provide information that is specific to that device type [8]. Intrusion Detection System (IDS) is a network device that aims at detecting all anomaly network attacks upon traffic analysis.

*C. Contribution and Paper Organization*

In [10], the authors proved that network anomalies can be detected using the SNMP-MIB and proper classifiers. They used a dataset with 34 variables categorized into five groups (Interface, IP, ICMP, TCP, and UDP). The motivation of this paper is to reduce the processes and operations performed by the IDS to learn the traffic patterns and classify the traffic. Our target is to increase the efficiency of the IDS by reducing the required resources. In this paper, we achieved the following points:

- Proved that using the SNMP-MIB Interface group variables and three machine learning classifiers (Lazy.IBk, Random Forest, and Random Tree), an accurate and effective detection model is designed with an accuracy rate of 99.94%.
- The SNMP-MIB Interface group variables are eight variables, instead of using 34 variables.

The rest of this paper is organized as follows: Section II discusses the related work in the area of network anomaly detection. The proposed model design and evaluation metrics are discussed in Section III. In Section IV, the experiment results are discussed and analyzed. Finally, the conclusion is illustrated in Section V.

## II. RELATED WORK

For more than two decades, researchers focused on anomaly detection because of the rapid increase of different network attacks. The anomaly detection can be useful in case of zero-day attacks. Some related anomaly detection approaches will be reviewed in this part.

In [10] [2], authors created a realistic dataset with 34 variables and 4998 records for network anomaly detection. This dataset was actually based on SNMP variables collected from network devices in a test-bed real network. The dataset includes 34 variables categorized in 5 groups (Interface, IP, ICMP, TCP, and UDP) and 8 network traffic categories (Normal traffic, TCP-SYN attack, UDP flood attack, ICMP-ECHO attack, HTTP flood attack, Slowloris attack, Slowpost attack, and Brute-Force attack). Many tools were used to generate real attacks like THC-Hydra 5.2, HyenaeFE, DOSHTTP 2.5.1, Slowloris script and ActivePerl language, and HttpDosTool4.0. Then, three machine learning classifiers were used: AdaboostM1, Random Forest, and MLP. Following experiment approach was stepped:

- A new SNMP-MIB dataset was created based on collected network traffic from a real test-bed environment.
- Accurate feature selection methods were used to select the most effective variables of the SNMP-MIB variables.
- Machine learning classification methods were applied to classify the attacks.
- The MIB variables were classified into 5 groups (Interface, IP, ICMP, TCP, and UDP).
- Each group was classified separately.

As a result, they found that attacks affect most on Interface, and IP groups rather than other groups. Another conclusion was that each group has a different classifier with high accuracy that may not give the same result with other groups.

In [11], another model was proposed to detect attacks anomaly using three machine learning techniques: BayesNet, Multi-Layer perception (MLP), and Support Vector Machine (SVM). The author used feature selection methods to increase model performance and reduce computation time. That was done using two feature selection methods with three attribute evaluators namely: InfoGain, ReleifF for Filter feature selection method, and GeneticSearch for Wrapper feature selection method. He introduced an acceptable detection accuracy model while reducing the complexity with 99.8% accuracy rate in 0.03 seconds.

Authors in [12] used a ready online repository dataset called Knowledge Discovery in Databases (KDD) that includes all types of attacks. In their work, they classified the attacks into 4 main categories that are DoS, U2R, R2L, and PROBE. They followed the following methodology:

- Performing KDD dataset pre-processing.
- Applying the prepared dataset in a fair environment.
- Examining all classifiers to find the most accurate one in detecting attacks.

They examined the following classifiers, Random forest, Multilayer Perception (MLP), Naïve Bayes, and Bayes Network. Finally, it was found that Random Forest classifier is the most accurate classifier to detect and classify the attacks in KDD dataset into their corresponding categories (DoS, U2R, R2L, and PROBE).

The model of [13] used a new feature selection model called Feature Vitality Based Reduction Method (FVBRM) based on three feature selectors that are Correlation –based Feature Selection (CFS), Information Gain (IG), and Gain Ratio (GR). FVBRM provided more accuracy than the three attribute selectors with 97.78% accuracy ratio.

In [14] [15], authors compared the performance of some machine learning classifiers (Support Vector Machine, Random Forest, Logistic Regression and Gaussian Naive Bayes), the comparison considered the following metrics (Accuracy, Precision, Recall, and F-Measure). The experiments were conducted on NSL-KDD dataset which is the enhanced version of KDD-cup 99. Among the tested classifiers, Random Forest performed the others with an accuracy rate of 99%.

In [16], authors evaluated the performance of 4 classifiers that are SVM, Naïve Bayes, Decision Tree and Random Forest using a big data processing tool network traffic intrusion detection called Apache Spark. Their evaluation metrics were based on accuracy, building time, and prediction time. The experiment was run on UNSW-NB15 dataset that has 42 features. It showed the superiority of the Random Forest classifier over the others in terms of accuracy, building time, and prediction time.

## III. PROPOSED MODEL

The proposed model performed the most accurate three machine learning classifiers on a ready-made SNMP-MIB Interface variables to measure the performance and compare among these classifiers. Waikato Environment for Knowledge Analysis (WEKA) 3.8 tool is used to conduct the experiment. The following sub-sections provide more details.

### A. SNMP-MIB Dataset

In this work, we adopted the SNMP-MIB dataset from [2] to perform the machine learning classification. This SNMP-MIB dataset is composed of 4998 records for network anomaly detection. This dataset was actually based on SNMP variables collected from network devices in a test-bed real network. The dataset includes 34 variables, which are categorized in 5 groups (Interface, IP, ICMP, TCP, and UDP) and 8 types of network traffic categories (Normal traffic, TCP-SYN attack, UDP flood attack, ICMP-ECHO attack, HTTP flood attack, Slowloris attack, Slowpost attack, and Brute-Force attack). Table I shows the number of records for each state of data traffic.

TABLE I. DATA TRAFFIC STATE RECORDS

| No | Traffic state (Normal/ Attack type) | Records |
|---|---|---|
| 1 | Normal | 600 |
| 2 | ICMP-ECHO Attack | 632 |
| 3 | TCP-SYN Attack | 960 |
| 4 | UDP Flood Attack | 773 |
| 5 | HTTP Flood Attack | 573 |
| 6 | Slowloris Attack | 780 |
| 7 | Slowpost Attack | 480 |
| 8 | Brute-Force Attack | 200 |

SNMP-MIB Dataset variables are categorized into five groups as in Table II.

TABLE II. SNMP-MIB VARIABLES AND CATEGORIES [2]

| Interface Group | IP Group | ICMP Group | TCP Group | UDP Group |
|---|---|---|---|---|
| ifInOctets | ipInReceives | icmpInMsgs | tcpOutRsts | udpInDatagrams |
| ifOutOctets | ipInDelivers | icmpInDestUnreachs | tcpInSegs | udpOutDatagrams |
| ifoutDiscards | ipOutRequests | icmpOutMsgs | tcpOutSegs | udpInErrors |
| ifInUcastPkts | ipOutDiscards | icmpOutDestUnreachs | tcpPassiveOpens | udpNoPorts |
| ifInNUcastPkts | ipInDiscards | icmpInEchos | tcpRetransSegs | |
| ifInDiscards | ipForwDatagrams | icmpOutEchoReps | tcpCurrEstab | |
| ifOutUcastPkts | ipOutNoRoutes | | tcpEstabResets | |
| ifOutNUcastPkts | ipInAddrErrors | | tcpActiveOpens | |

Our proposed model depends on SNMP-MIB interface variables to detect network anomalies. In Table III, a brief description of each interface variable.

TABLE III. INTERFACE VARIABLE DISCRIPTION

| Label | Variable Name | Variable description |
|---|---|---|
| 1 | ifInOctets | The total number of octets received on the interface. |
| 2 | ifOutOctets | The total number of octets transmitted out of the interface |
| 3 | ifoutDiscards | The number of outbound packets which were chosen to be discarded even though no errors had been detected to prevent their being transmitted |
| 4 | ifInUcastPkts | The number of packets, delivered by this sub-layer to a higher (sub-) layer, which is not addressed to a multicast or broadcast address at this sub-layer. |
| 5 | ifInNUcastPkts | The number of packets, delivered by this sub-layer to a higher (sub-)layer, which is addressed to a multicast or broadcast address at this sub-layer |
| 6 | ifInDiscards | The number of inbound packets which are chosen to be discarded even though no errors had been detected to prevent them from being deliverable to a higher-layer protocol |
| 7 | ifOutUcastPkts | The total number of packets that higher-level protocols requested be transmitted, and which were not addressed to a multicast or broadcast address at this sub-layer |
| 8 | ifOutNUcastPkts | The total number of packets that higher-level protocols requested be transmitted, and which were addressed to a multicast or broadcast address at this sub-layer, including those that were discarded or not sent |

### B. Machine Learning Classifiers

Machine learning classifiers are algorithms that understand the dataset and make predictions [17]. The classifier uses the SNMP-MIB dataset records to train and build a classification model. Using the classifier, the model can classify new objects with high accuracy. As mentioned before, the top three classifiers are used in building the model. These classifiers are illustrated as follows.

Lazy.IBk machine learning algorithm stores the training instances and does nothing until classification is triggered. Lazy learning is a learning method in which generalization beyond the training data is delayed until a query is made to the system where the system tries to generalize the training data before receiving queries [18].

Random Committee classifier builds an ensemble of base classifiers and averages their predictions. Each one is based on the same data but uses a different random number of seeds [19].

Random Forest classifier was introduced in 2000's by Leo Breiman that combines a set of decision trees that grow by selecting random subspaces of the dataset [20].

## C. Feature Selection

Feature selection is a method to find the most related features (attributes) that may increase the accuracy and the efficiency of the classification with a minimal amount of hardware resource consumption in terms of memory, CPU, and bandwidth.

In this model, all attribute evaluators provided by the WEKA tool are implemented on the interface group variables. The first top 5, and 3 results of each evaluator are taken as input to the top 3 classifiers to classify the dataset records again based on these 5, and 3 features.

## D. Evaluation Metrics

In this paper, the most well-known measurement metrics had been adopted namely, Precision, Recall, and F-Measure.

Precision is a ratio of correctly predicted positive samples to all predicted positive samples, where Recall is the ratio of correctly predicted positive samples to the total number of real positive samples. F-Measure takes in its consideration the two previous metrics as in the following formulas:

$$\text{Precision} = \frac{TP}{TP + FP} \quad \ldots\ldots\ldots(1)$$

$$\text{Recall} = \frac{TP}{TP + FN} \quad \ldots\ldots\ldots(2)$$

$$F\text{-Measure} = 2\frac{Precision \cdot Recall}{Precision + Recall} \quad \ldots\ldots\ldots(3)$$

## IV. EXPERIMENTAL RESULTS

In this experiment, all classifiers (43 classifiers) were applied to the interface variables dataset to choose the most three accurate classifiers. From Table IV, the most three accurate classifiers were Lazy.IBk (LI), meta.RandomCommittee (RC), and trees.RandomForest (RF).

TABLE IV. CLASSIFIERS RESULTS

| No. | Classifier | Correctly Classified Instances Rate (%) |
|---|---|---|
| 1 | lazy.IBk | 99.94 |
| 2 | meta.RandomCommittee | 99.9 |
| 3 | trees.RandomForest | 99.9 |
| 4 | meta.RandomSubSpace | 99.8599 |
| 5 | trees.RandomTree | 99.8399 |
| 6 | rules.PART | 99.6999 |
| 7 | meta.ClassificationViaRegression | 99.6799 |
| 8 | meta.AttributeSelectedClassifier | 99.6399 |
| 9 | trees.J48 | 99.5998 |
| 10 | meta.Bagging | 99.5798 |
| 11 | trees.REPTree | 99.4798 |
| 12 | bayes.BayesNet | 99.3798 |
| 13 | meta.FilteredClassifier | 99.3597 |
| 14 | rules.JRip | 99.2797 |
| 15 | trees.LMT | 98.0392 |
| 16 | meta.IterativeClassifierOptimizer | 97.8191 |
| 17 | meta.LogitBoost | 97.8191 |
| 18 | functions.MultilayerPerceptron | 97.2789 |
| 19 | rules.DecisionTable | 96.9588 |

Figs. 1 and 2 show the performance results of F-Measure and Precision of the three classifiers on all interface variables dataset. All three classifiers performed 100% in classifying Slowpost and Brute force traffic. Lazy.IBk performed better than others in classifying normal traffic.

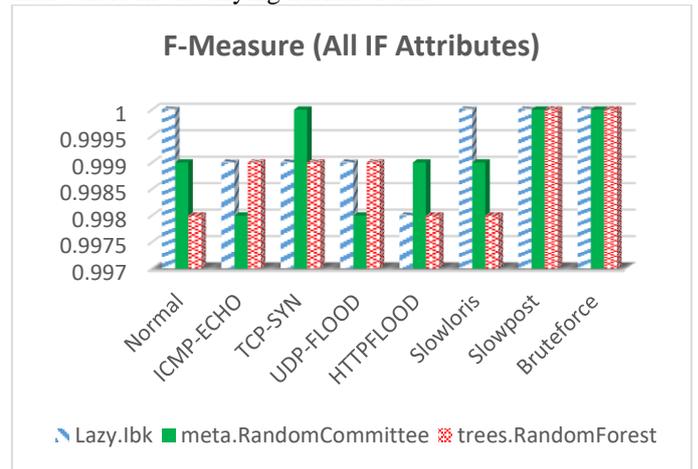

Fig. 1: F-Measure Results for Adopted Classifiers

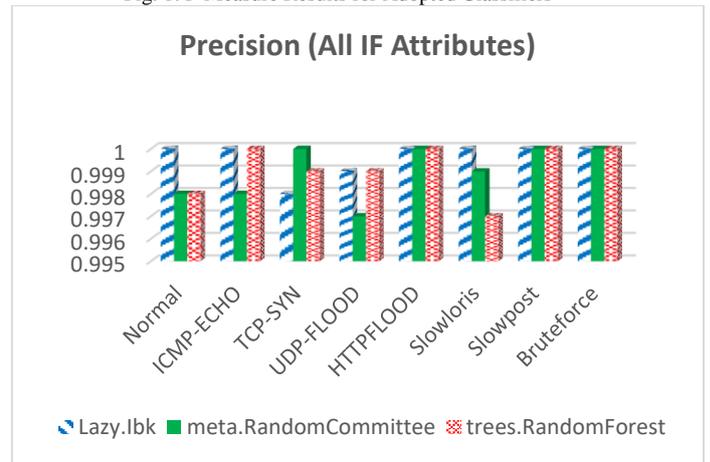

Fig. 2: Precision Results for Adopted Classifiers

In this paper, all attribute evaluators are implemented on the interface group variables as in Table V. The first top 5, and 3 results of each evaluator are taken as input to the top 3 classifiers to classify attacks again based on these 5, and 3 attributes. High accurate results in classification mean a good attribute evaluation. The most accurate attribute evaluators are InfoGain, ReliefF, and Correlation, when used with Lazy.IBk classifier with a rate of 100% Correctly Classified Instances. InfoGain, ReliefF, and Correlation attribute evaluators had selected the same attributes that are 2, 4, 5, 7, and 8.

TABLE V. ATTRIBUTE EVALUATORS RESULTS

| No | Attribute Evaluator | Top 5 attributes | Classifier | Correctly Classified Instances Rate | |
|---|---|---|---|---|---|
| | | | | Top 5 | Top 3 |
| 1 | InfoGain | 8,7,4,5,2 | LI | 100% | 99.98 |
| 2 | InfoGain | 8,7,4,5, 2 | RC | 99.9% | 99.92 |
| 3 | InfoGain | 8,7, 4,5,2 | RF | 99.94% | 99.96 |
| 4 | Correlation | 4,8,5,7,2 | LI | 100% | 100 |
| 5 | Correlation | 4,8,5,7,2 | RC | 99.9% | 99.88 |
| 6 | Correlation | 4,8,5,7,2 | RF | 99.94% | 99.94 |
| 7 | CfsSubsetEval | 2,3,4,7,8 | LI | 99.98% | 99.4798 |
| 8 | CfsSubsetEval | 2,3,4,7,8 | RC | 99.88% | 98.4994 |
| 9 | CfsSubsetEval | 2,3,4,7,8 | RF | 99.92% | 98.8996 |
| 10 | ReliefF | 5,8,4,7,2 | LI | 100% | 100 |
| 11 | ReliefF | 5,8,4,7,2 | RC | 99.9% | 99.88 |
| 12 | ReliefF | 5,8,4,7,2 | RF | 99.94% | 99.94 |
| 13 | ClassifierAttributeEval | 8,7,2,3,4 | LI | 99.98% | 99.96 |
| 14 | ClassifierAttributeEval | 8,7,2,3,4 | RC | 99.88% | 99.94 |
| 15 | ClassifierAttributeEval | 8,7,2,3,4 | RF | 99.92% | 99.96 |

The following Figs. 3 and 4 show the performance results (F-Measure and Precision) for the three classifiers on top 5 interface attributes.

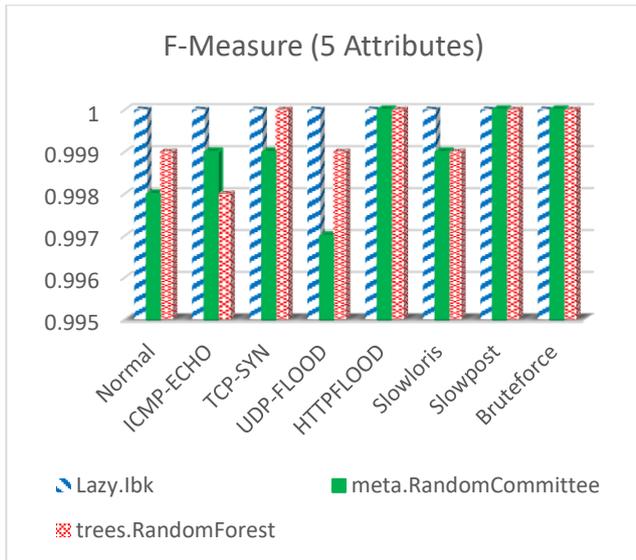

Fig. 3:F-Measure Results (Top 5 Attributes)

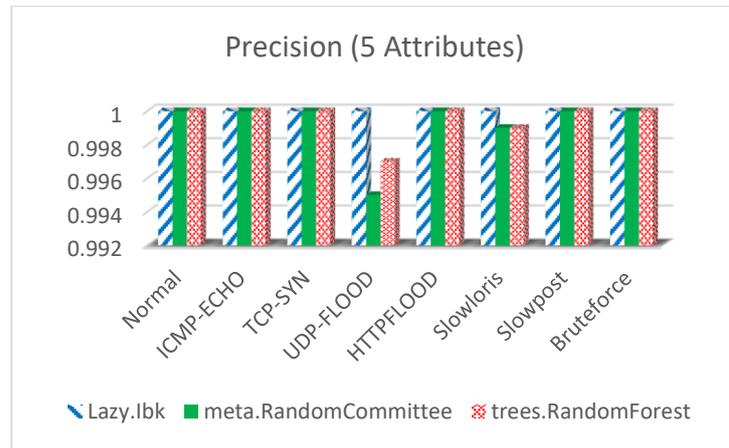

Fig. 4:Precision Results (Top 5 Attributes)

As in Table 5, the best correctly classified instances rate is for InfoGain, Correlation, and ReliefF attribute evaluators with Lazy.IBk classifier with result 99.98%, 100%, and 100%, respectively. InfoGain selected the attributes 8, 7, and 4 as most related attributes. Both Correlation, and ReliefF selected the attributes 5, 8, and 4 as most related attributes. The results in Fig. 5, 6, 7, and 8 indicate that Lazy.IBk classifier achieved high performance (100%) with most three related attributes in classifying all attacks in the dataset. Another point is that attributes selected from Correlation, and ReliefF has given higher precision, and F-Measure results in more than attributes selected by InfoGain.

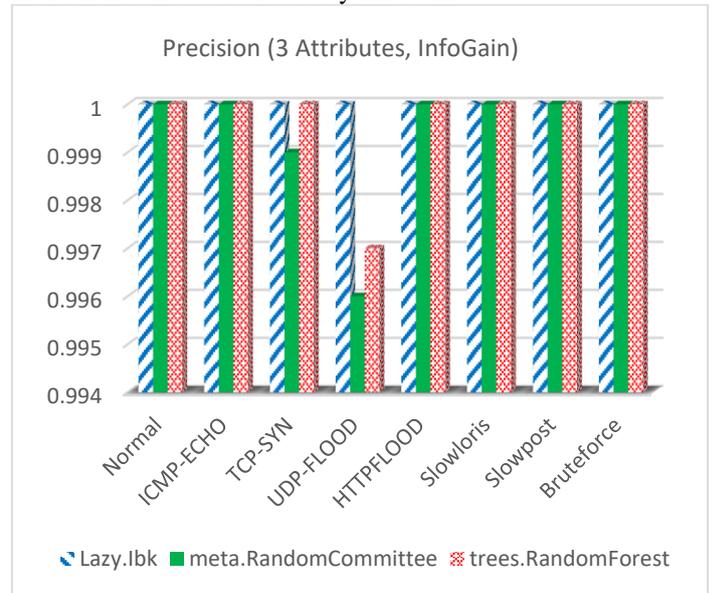

Fig. 5: Precision Results (3 Attributes by InfoGain)

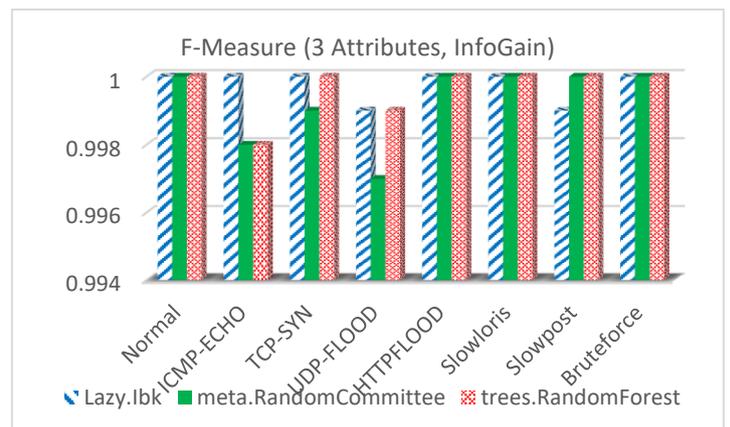

Fig. 6:F-Measure Results (3 Attributes by InfoGain)

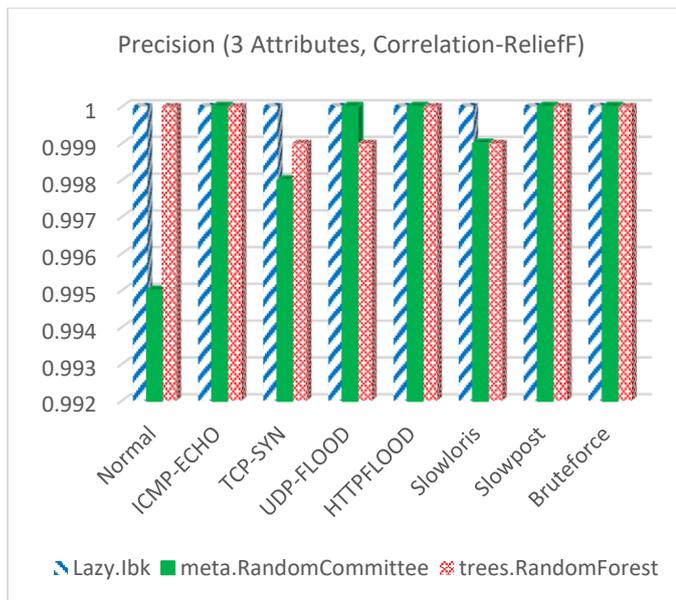

Fig. 7: Precision (3 Attributes by Correlator and ReleifF)

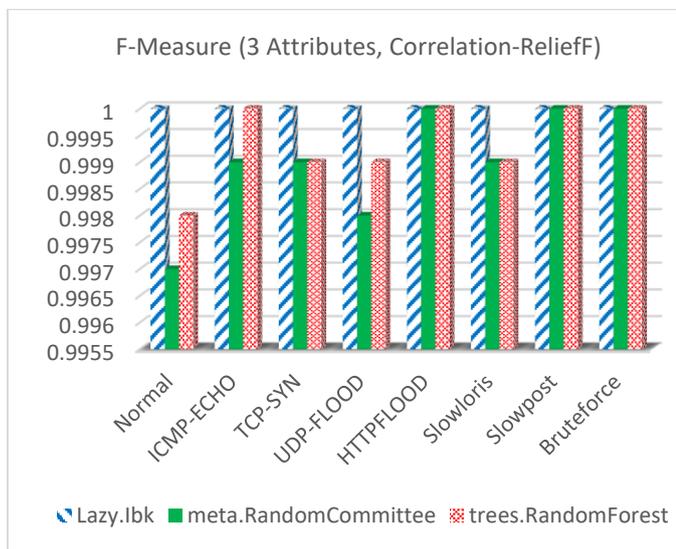

Fig. 8: F-Measure (3 Attributes by Correlator and ReleifF)

## V. CONCLUSIONS

In this paper, SNMP-MIB interface features were exploited to detect brute-force and DoS attacks. Machine learning algorithms were used to learn dataset records and to predict the matching attack. There was a variance in machine learning classifiers' performance in terms of precision and F-Measure metrics. The most appropriate classifiers to classify attacks using interface variables were Lazy.IBk, meta.RandomCommittee, and trees.RandomForest that gave high accurate results. Attribute evaluators were used to selecting the most effective 5, and 3 attributes to minimize the number of attributes processed by the classifier, and then to reduce the allocated resources in real IDS devices to detect the anomalies. The most accurate attribute evaluators were Correlator, and ReliefF, that when used with Lazy.IBk classifier on interface parameters gave 100% accurate results.

Finally, applying Correlator, and ReliefF attribute evaluators, and Lazy.IBk machine learning classifier in IDS increased the performance, and accuracy of IDS functionality with minimum hardware resources consumption (CPU, memory, and bandwidth).